# In-situ imaging of the three-dimensional shape of soft responsive particles at fluid interfaces by atomic force microscopy

Jacopo Vialetto[a],* Shivaprakash N. Ramakrishna[a],* and Lucio Isa*

*Laboratory for Soft Materials and Interfaces, Department of Materials, ETH Zürich,
Vladimir-Prelog-Weg 5, 8093 Zürich, Switzerland*

E-mail: jacopo.vialetto@mat.ethz.ch; shivaprakash.ramakrishna@mat.ethz.ch;
lucio.isa@mat.ethz.ch

---
[a]These two authors contributed equally.

**Abstract**

The reconfiguration of soft, deformable particles upon adsorption at the interface between two fluids underpins many aspects of their dynamics and interactions, ultimately controlling the macroscopic properties of particle monolayers of relevance for materials, such as particle-stabilized emulsions and foams, and processes, e. g. particle-based lithography. In spite of its importance, experimentally determining the three-dimensional shape of soft particles at fluid interfaces with high resolution remains an elusive task. In this work, we take poly(N-isopropylacrylamide) (pNIPAM) microgels as model soft particles and demonstrate that their conformation at the interface between an aqueous and an oil phase can be fully reconstructed by means of *in-situ* atomic force microscopy (AFM) imaging. We show that imaging the particle topography from both sides of the interface allows one to characterize the in-plane deformation of the particle under the action of interfacial tension and to visualize the occurrence of asymmetric swelling in the two fluids. Additionally, the technique enables investigating different fluid phases and particle architectures, as well as studying *in situ* the effect of temperature variations on particle conformation. We envisage that these results open up an exciting range of possibilities to provide microscopic insights between the single-particle behavior of soft objects at fluid interfaces and macroscopic material properties of relevance for applications and fundamental studies alike.

## INTRODUCTION

The confinement of colloidal particles at fluid interfaces holds the key for a broad range of phenomena with applied and fundamental relevance alike, including the stabilization of emulsions and foams,[1,2] the encapsulation and manipulation of liquids[3,4] and the creation of model two-dimensional (2D) materials.[5–9]

In the case of hard, mechanically rigid particles of a given shape, all aspects of their adsorption/desorption, dynamics and interactions with and at the interface are influenced by a single parameter, the particle contact angle $\theta$, which defines the position of the particle with respect to the interface plane.[10] Due to its fundamental importance, many techniques have been developed to measure $\theta$.[11,12] However, if the particle is deformable, it can reconfigure upon adsorption under the action of interfacial tension and due to exposure to different solvents.[13,14] Conformational changes and anisotropic deformations relative to the bulk imply that particle properties at the interface can no longer be ascribed to a single parameter and that the notion of a contact angle may no longer be well defined. This more complex response is closely connected to the emergence of additional properties and functionalities, which make soft particles at fluid interfaces highly interesting in formulations, as platforms for materials fabrication, and for more fundamental understanding on the 2D phase behavior of compressible objects.[15,16] Consequently, new experimental approaches are required to characterize the three-dimensional (3D) shape of soft particles adsorbed at fluid interfaces and infer how this affects their adsorption and desorption, dynamics and interactions.

Among a broad class of colloidal-scale objects, microgels, i.e. crosslinked polymer particles swollen by the solvent in which they are dispersed, have emerged as a powerful and versatile system. The ease and multiplicity of synthetic strategies to obtain microgels with different internal architectures and polymer compositions[17] makes them ideal model systems to elucidate how these parameters affect the adsorption and organization of soft objects at fluid interfaces.[16] This has allowed their use as synthetic counterparts to complex proteins and bio-polymeric colloids[18] and as promising elements for the realization of complex 2D materials.[19,20] Moreover, the incorporation of stimulus-responsive (e.g. temperature, pH, light, etc.) polymers identifies microgels as key elements in smart formulations.[21] However, as a consequence of their relatively small size and low refractive index mismatch with the solvents, accessing detailed microscopic information on their conformation at the interface remains a daunting task.

Most characterization techniques with single-particle resolution either rely on *ex-situ* investigations, i.e. after transferring the particles from the interface onto a solid support, or give incomplete data, e.g. can only visualize the particle shape with insufficient resolution or have access to one side of the interface only. In particular, *in-situ* techniques based on electron microscopy, such as cryo-SEM[22] and freeze-fracture shadow-casting (FreSCa) cryo-SEM,[23,24] or transmission X-ray microscopy,[25] require fast freezing of the samples and cannot be used to probe the particle response to stimuli in ambient conditions. Moreover, the first two only expose one side of the microgels at the interface and cannot provide real 3D reconstructions. Optical microscopy, including confocal microscopy, requires the use of fluorescent markers and, in any case, does not provide sufficiently high spatial resolution.[22,26,27] Conversely, complementary approaches for *in-situ* characterization, such as ellipsometry[28] or neutron reflectivity,[29] provide accurate information on the thickness of adsorbed microgel layers but rely on strong assumptions to extract single-particle conformation. Finally, even if deposition and *ex-situ* analysis has been an extremely valuable tool for characterizing both the single-particle properties[24,30–33] and the microstructure of the resulting monolayers,[20,34] it has some limitations. The presence of specific interactions between particles and the substrate used may affect the transfer of the microgels and their resulting conformation.[35] But, most importantly, the technique can only resolve a 2D projection of the polymer density distribution across the interface for a particle in a dry state and does not give direct access to its 3D conformation at the interface.

In this work, we propose an alternative approach that enables imaging the full 3D shape of soft particles adsorbed at oil-water interfaces at high resolution, using poly(N-isopropylacrylamide) (pNIPAM) microgels as model systems, by using *in-situ* atomic force microscopy (AFM).

*In-situ* AFM imaging at fluid interfaces has been previously applied to closely-packed nanoparticle monolayers[36–38] and polymeric films,[39,40] to capture their microstructure in real space with exceptionally high lateral and vertical resolution. Here, we greatly extend the applicability of this technique to include: i) the imaging of dilute layers of soft polymeric particles, allowing us to disclose the reconfiguration of the polymer network upon adsorption at the fluid interface at the single-particle level; ii) complementary imaging from the oil and water phases to obtain a full 3D shape reconstruction with nanometric resolution; iii) temperature-resolved imaging to monitor *in-situ* the response of the pNIPAM network on both sides of the interface below and above the volume phase transition temperature (VPTT) of the microgels (T $\sim$ 32°C in water). The versatility of the technique also allows investigating other system parameters, e. g. different organic phases, unraveling how the conformation of the adsorbed particles adapts to changes in the interfacial tension and in the partial solubility of the polymer in the two fluids. Moreover, by using microgels with different internal architecture (in terms of crosslinker content and distribution), we demonstrate that particle design directly affects their conformation at the interface, and is thus a crucial parameter influencing the structural and mechanical properties of microgel monolayers.

# RESULTS

## Three-dimensional AFM imaging of isolated particles

We begin by describing the experimental setup and the capabilities of the method. Fig. 1 schematically illustrates the measurement conditions used for AFM imaging at a liquid-liquid interface. In particular, we use two configurations in order to image adsorbed particles from *both sides* of the fluid interface. In a first set of experiments, a dilute aqueous microgel suspension is confined within a thin ring made of UV-curable glue on a silicon wafer (Fig. S1). Subsequently, the cell containing the silicon wafer is filled with hexadecane to form the fluid interface to which particles spontaneously adsorb by diffusion. After an equilibration time of about 30 minutes, the tip is approached to the interface from the oil side, and AFM images are acquired by means of PeakForce tapping-mode (Fig. 1a). In a complementary set of experiments, the thin ring on the silicon wafer is instead filled with hexadecane. A drop of an aqueous suspension of the microgels is then placed on top of the ring to form the interface, and the system is left to equilibrate for approximately 5 minutes, during which particles reach the fluid interface and adsorb there. Subsequently, the cell is filled with water to remove excess particles not yet adsorbed to the fluid interface. After an additional equilibration time of about 30 minutes, we then approach the interface with the AFM tip from the water side and acquire PeakForce tapping-mode images (Fig. 1c).

The combination of these two imaging configurations allows for the *in-situ* capturing of the complex 3D conformation of adsorbed soft particles virtually in the same experimental conditions. Representative AFM images of monolayers of standard pNIPAM microgels (labeled *CC5*, showing the typical core-corona profile in bulk water, with 5 mol % BIS cross-linker and hydrodynamic diameter $D_h = 1150 \pm 27$ nm, see Methods) are reported in Fig. 1b (oil side) and Fig. 1d (water side). The technique nicely captures the ordered hexagonal arrangement of the particles in the monolayer, from both fluid phases.

From such images, we can also extract quantitative information on the protrusion profiles of the microgels in both phases and on the polymer distribution within the interface plane. Imaging from hexadecane reveals that the microgel is collapsed and barely protrudes out of the interface, reaching, for these particles, a maximum height of 71 ± 5 nm (Fig. 1e). The thickness of the polymer layer decreases from the center towards the edge of the particle, and stretches on the plane of the interface for approximately 500 nm in radius. At larger distances from the particle center, the polymer chains adsorbed on the fluid interfaces are no longer detectable from the height images, while they may remain visible in the adhesion images (see Fig. S3).

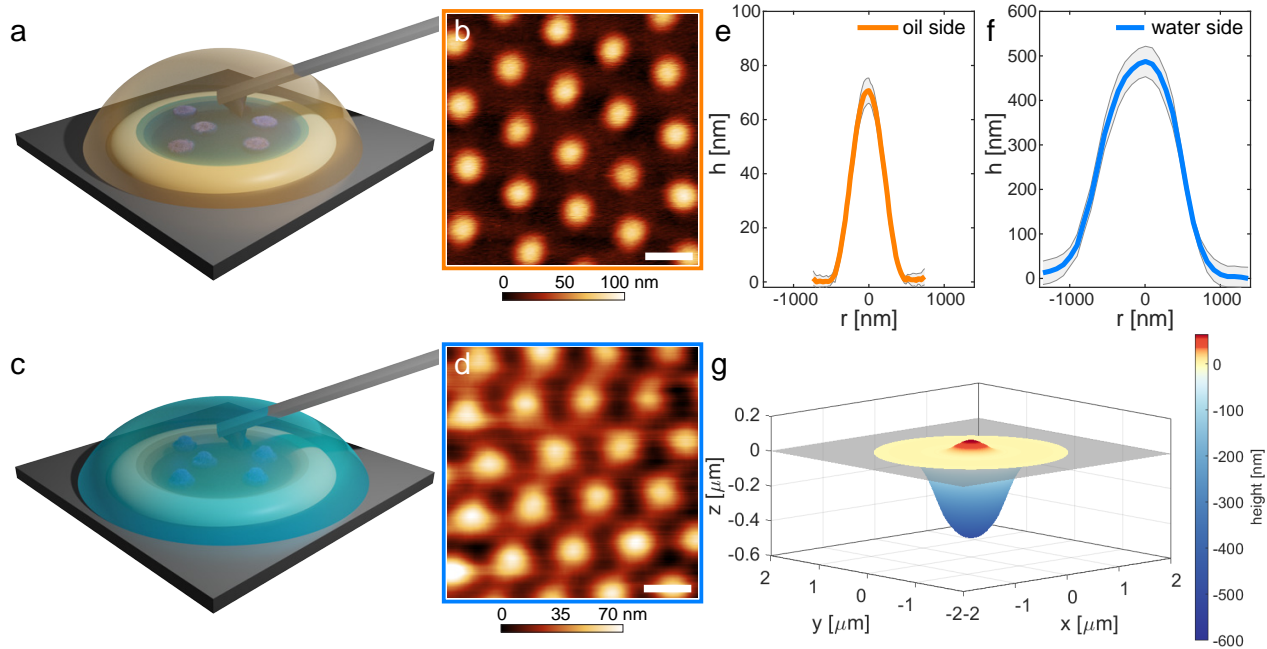

Figure 1: **3D imaging of microgels adsorbed at a hexadecane-water interface.** (a) Sketch of the measurement configuration for AFM imaging at the interface between water (subphase) and hexadecane (top phase). (b) AFM height image of a microgel monolayer visualized from the oil side. (c) Sketch of the complementary measurement configuration with hexadecane as the subphase and water as the top phase. (d) AFM height image of a microgel monolayer visualized from the water side. Scale bars: 1 $\mu$m. (e-f) Mean height profiles of adsorbed microgels imaged from the oil (e) and the water (f) side, respectively (corresponding AFM images in Fig. S2). The shaded regions correspond to the standard deviations of the height profiles calculated on at least 10 particles. (g) Reconstructed 3D profile across the interface. The gray rectangle indicates the interface plane.

The complementary images from the water side show a significantly different height profile, which is strongly influenced by the packing of the microgels. In particular, the swelling of the polymer network in water implies that the full range of the height profiles can only be detected for isolated or well-separated particles, where the interface plane is also visible as a reference (Fig.

S2). Similarly to the profiles from the oil side, the polymer content decreases from the center of the particle toward the edge. However, the peak height is much greater (490 ± 30 nm) and the in-plane dimensions extend to approximately 2100 ± 250 nm.

Merging these two height profiles allows for a complete 3D reconstruction of the conformation of the microgel adsorbed at an oil-water interface, as reported in Fig. 1g. The resulting profile matches the finding of asymmetric shapes deduced by FreSCa cryo-SEM experiments, numerical simulations, and AFM images of microgels transferred onto a solid substrate,[23,24,33] evidencing what has been called as a "fried-egg" shape. However, our measurements provide a direct, quantitative description, which escaped previous approaches.

## Imaging particles in contact

After determining the shape of isolated particles, we now move to investigating the conformation of adsorbed microgels in contact. In Figs. 2a-b we report AFM height images of two neighbouring particles from both sides of the interface. The reconstructed profiles (Fig. 2c) show that, at the same center-to-center separation, there is significant overlap between the two polymer networks in the water phase, forming a large contact region below the interface with possible compression and interpenetration of the outer part of the microgels, even in the absence of external compression of the interface. Conversely, from the oil side, the particles only sterically interact through their outermost polymer chains adsorbed onto the plane of the interface. The detected height profiles essentially decay to zero in the contact region and the presence of interacting chains is only visible in the adhesion images at this magnification (Fig. S3a).

More insights can however be gained at higher magnification as shown in Figs. 2d-f. The close-up view of a compact monolayer imaged from the oil side illustrates that high-resolution height images (e.g. as in Fig. 2d) nicely capture the collapsed chains on the surface of the microgels exposed to hexadecane, which aggregate forming globules and bundles onto the particle core, similar to the conformation of collapsed pNIPAM chains measured at high temperature in aqueous conditions on solid supports.[41] The image also shows that the polymer corona appears to be preferentially localized in the contact regions, forming "polymer bridges". Even if the extent of interpenetration between the coronae of neighbouring particles is not directly measurable from the images, examining the adhesion (Fig. 2e) and deformation (Fig. 2f) channels clearly shows that the particles deform and compress into a closely packed honeycomb contact network.

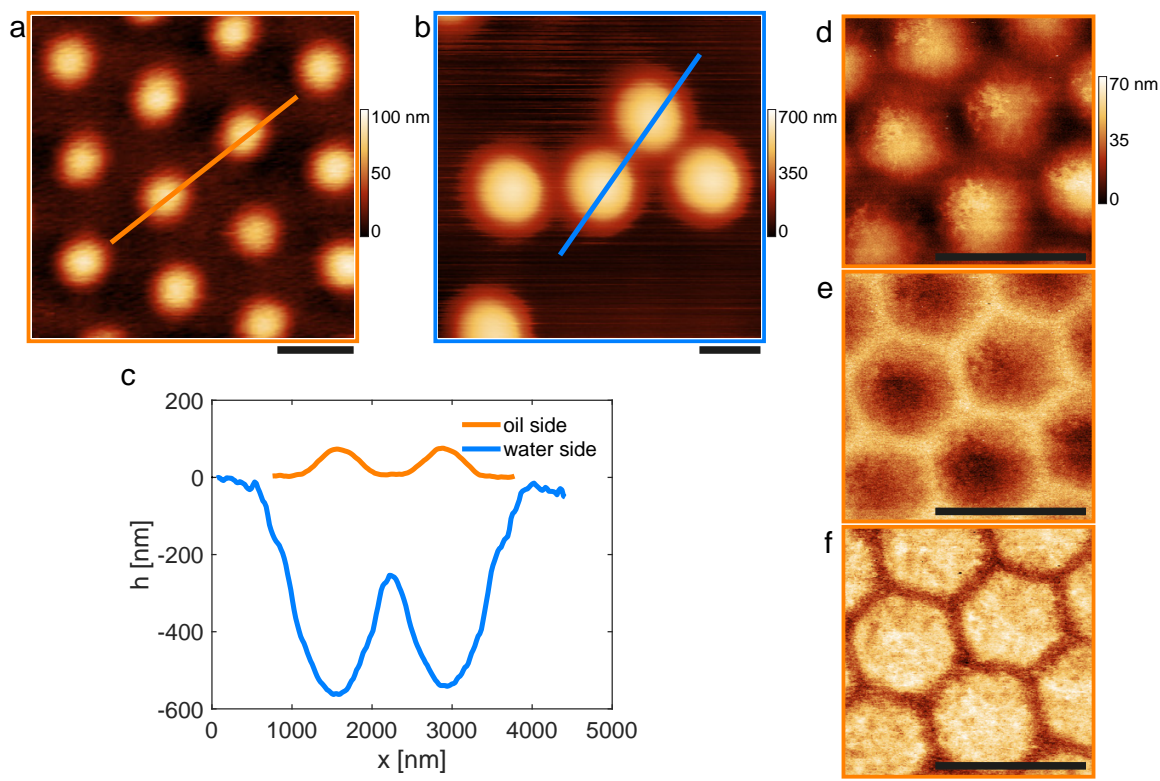

Figure 2: **Conformation of microgels in contact.** (a-b) AFM height images of *CC5* microgels at the hexadecane-water interface visualized from the oil (a) and water (b) side, respectively. (c) Height profiles extracted from the images in (a-b) along the indicated lines. (d-f) AFM images of a close-packed monolayer visualized from the oil side. (d) height, (e) adhesion and (f) deformation image. Scale bars for all images: 1 $\mu$m.

These results undoubtedly indicate that the interactions among adsorbed microgels occur both on the plane of the interface, through the adsorbed and stretched polymer polymer coronas, and in the good solvent, where the peripheries of the particles overlap.

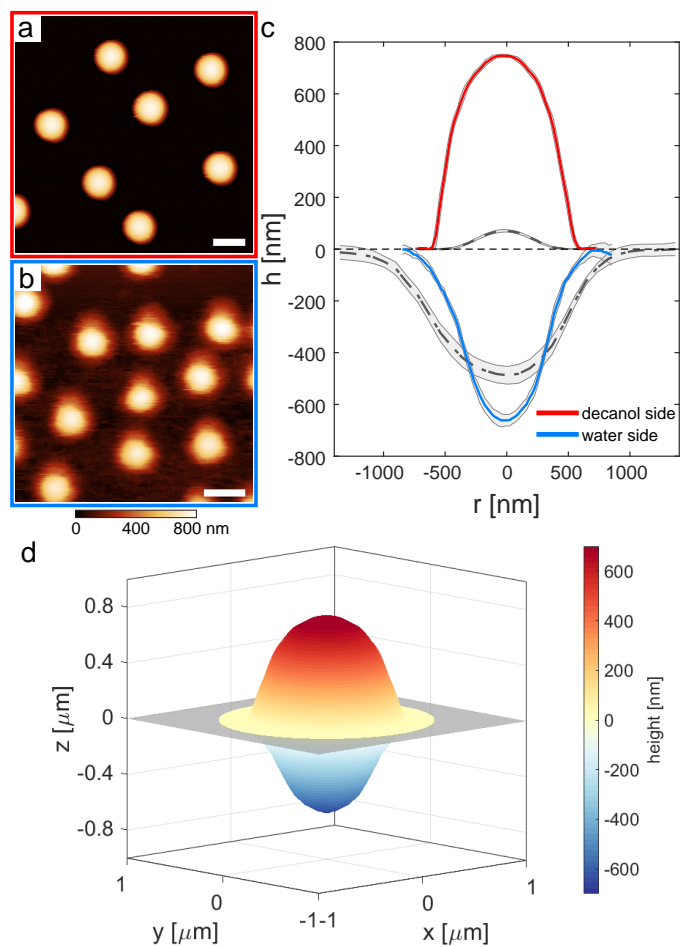

Figure 3: **Influence of the organic phase on the 3D conformation of adsorbed microgels.** (a-b) AFM height images of *CC5* microgels adsorbed at a 1-decanol-water interface, imaged from the 1-decanol (a) and water (b) side, respectively. Scale bars: 1 $\mu$m. (c) Averaged height profiles of the microgels at the 1-decanol-water interface. The grey dashed and dash-dotted lines represent profiles of the same microgels at the hexadecane-water interface, from the hexadecane and water side, respectively. (d) Reconstructed 3D profile across the interface. The gray rectangle indicates the interface plane.

## Effect of the oil phase

The findings reported above are typical for the case of conventional core-corona microgels exposed to a non-polar oil with high interfacial tension and where pNIPAM is poorly soluble. Our approach nonetheless enables us to probe the influence of both polymer solubility and interfacial tension on the 3D conformation of the adsorbed microgels by imaging through different oils. In particular, we expect the interfacial tension ($\gamma$) to dictate the microgel deformation within the interface plane, and to define how the polymer network rearranges upon lateral compression.[42] In order to examine markedly different cases, we replaced hexadecane by 1-decanol, therefore switching $\gamma$ values from $\simeq 50 mN \cdot m^{-1}$ for the hexadecane-water system to $\simeq 9 mN \cdot m^{-1}$ for the 1-decanol-water system. In addition to the drop in interfacial tension, pNIPAM is soluble in fatty alcohols and consequently the microgels are expected to swell both in the water and in the oil phase.[43]

This hypothesis is confirmed by the *in-situ* AFM height images, reported in Fig. 3a-b, and by the corresponding height profiles (Fig. 3c). For a direct comparison, the profiles of the same particles at an hexadecane-water interface are also shown in Fig. 3c (grey dash and dash-dotted lines). The particles at the 1-decanol-water interface show a similar degree of swelling on both sides of the interface, resulting in an almost symmetrical shape (see also the reconstructed 3D profile in Fig. 3d), which is qualitatively different from the highly asymmetric 3D conformation of the same particles at the hexadecane-water interface. Our imaging enables the quantification of the position of the particle relative to the interface plane, measured as the height ratio $h_w/h_o$, where $h_w$ and $h_o$ are the maximum height of the particle in water and oil, respectively. The more homogeneous swelling at the 1-decanol-water interface gives $h_w/h_o = 0.89 \pm 0.04$, as opposed to the highly asymmetric conformation of the microgels at the hexadecane-water interface, for which $h_w/h_o = 6.9 \pm 0.9$. This quantification clearly evidences the effect of the solubility in the organic phase on the rearrangement of soft polymeric particles.

Moreover, the reduced value of the interfacial tension leads to a lower deformation within the interface plane, with an interfacial diameter (measured from the water side) of $D_i \simeq 1520$ nm at the 1-decanol-water interface relative to a value of 2100 nm for the hexadecane-water system. The particle diameter at the interface ($D_i$) can be used to quantify the stretching ratio of the particles at the interface with respect to their spherical shape in bulk aqueous conditions, defined as $D_i/D_h$, where $D_h$ is the hydrodynamic diameter measured by dynamic light scattering. The calculated stretching ratio decreases from $1.8 \pm 0.3$ to $1.3 \pm 0.2$, from hexadecane to 1-decanol, indicating the lower degree of deformation at the 1-decanol-water interface. The deviation from a spherical shape is also primarily concentrated in proximity of the interface plane (Fig. 3d), similar to prediction for neutrally wetting soft spheres.[13]

Notably, when imaging denser monolayers, we observe that closely-packed hexagonal microgel assemblies can also be obtained at the 1-decanol-water interface (Fig. S4). The high swelling on both sides and the limited in-plane deformation prevent accessing the conformation of the polymer corona at the interface, and the particles appear to retain an isotropic shape without rearranging into facets as it was evidenced at the hexadecane-water interface by imaging the collapsed particles through the oil phase. For this case of symmetrical high swelling, the description of the shape of the microgels at the interface as "fried eggs" is no longer applicable.

## Effect of the particle architecture

So far we have examined only one particle type. However, the complex 3D conformation of a soft particle adsorbed at fluid interface is intimately related with the internal architecture of its polymer network, i.e. resulting from the synthesis protocols used.[33] In Fig. 4 we report a detailed quantification of the profiles of three different microgels adsorbed at the hexadecane-water interface, as a function of their internal polymer density profiles in bulk as measured by static light scattering (Figs. 4a-c; see Methods). Microgels *CC5* and *CC1* have the typical core-corona profile in water at 25°C, with a denser core and a decrease in polymer content towards the periphery of the particles.[44,45] They differ by the crosslinking content, which is 5 and 1 mol % BIS for *CC5* and *CC1*, respectively, allowing to investigate the effect of the particle internal elasticity on the network deformation upon interfacial adsorption. Microgel *INV* is instead obtained *via* a two-step synthesis process (see Methods), which confers an "inverse" polymer density profile, with a less-dense core and a more crosslinked shell. It is therefore characterized by a qualitatively different density profile than that of *CC5* and *CC1*.

We first describe our observations concerning the shape the particles assume at the fluid interface. We then characterize their full 3D conformation after adsorption by the height ($h_w/h_o$) and swelling ($D_i/D_h$) ratios.

The two core-corona microgels have a similar 3D profile after adsorption to the fluid interface (Figs. 4d, e), with the denser core that protrudes more into the water phase, and with a polymer content that continuously decreases towards the particle periphery. The microgel *INV* is, instead, characterized by a much flatter profile on the oil side, with an almost constant thickness up to the visible particle periphery (Fig. 4f). Its conformation in the water phase resembles that of *CC5* and *CC1*, however with quantitative differences, as detailed below. The height ratio increases from $6.9 \pm 0.9$ for *CC5*, to $13.8 \pm 1.6$ and $33 \pm 8$ for *CC1* and *INV*, respectively. In the case of more cross-linked particles, their decreased deformability leads to the protrusion of a higher amount of polymer in the organic, immiscible phase. Conversely, for decreasing the cross-linking density, and even more so in the absence of a cross-linked core, the particle can stretch further on the interface plane. This characteristic is quantified by the stretching ratio, which for microgel *CC5* is equal to 1.8 ± 0.3, while it reaches 2.0 ± 0.3 for *CC1*, in agreement with previous studies reporting an increase in the particle elongation on the interface plane for less cross-linked, and therefore softer, microgels.[24] The absence of a cross-linked core allows the *INV* particle to deform even more at the interface in order to maximize the amount of adsorbed polymer, reaching a value of $D_i/D_h = 2.4 ± 0.3$.

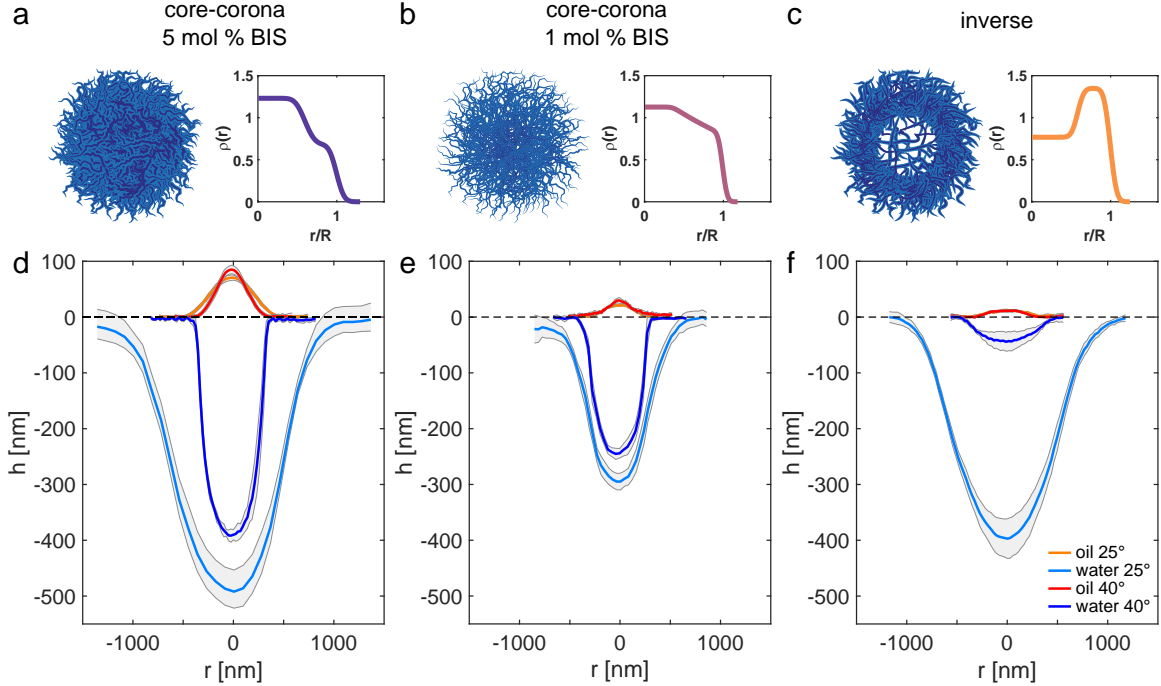

Figure 4: **Conformation of individual microgels at a hexadecane-water interface as a function of internal architecture and temperature.** (a-c) Sketch the microgels' internal architecture and corresponding polymer density profile ($\rho(r)$) at 25°C plotted as a function of a normalized radial coordinate $r/R$, where $R$ is the particle radius in bulk, as extracted from fitting of the static light scattering form factors (see Methods). (d-f) Height profiles from AFM images taken from the oil (orange and red curves for profiles at 25° and 40°C, respectively) and water side (light and dark blue curves for profiles at 25° and 40°C, respectively). The shaded regions correspond to the standard deviations of the height profiles calculated on at least 10 particles. The dotted line indicates the interface plane. The corresponding AFM images are in Figs. S2, S5, S6. (a,d) Core-corona microgel with 5 mol % BIS. (b,e) Core-corona microgel with 1 mol % BIS. (c,f) Microgel with an inverse profile, ultra-low crosslinked core and 5 mol % BIS in the shell.

To conclude, it is instructive to compare *in-situ* AFM imaging with the profiles of the same microgels after transfer on a solid substrate and imaging in the dried state (Fig. S7), as this is a commonly used technique to infer information over the microgels conformation at the fluid interface.[9,24] The "dry" height profiles in Fig. S7d show that the overall shape of *CC5* and *CC1* microgels is qualitatively captured, with a Gaussian-like profile that resembles the one imaged from the water side. Conversely, the "dry" height profile of the *INV* microgel does not match the shape that the particles had at the fluid interface, emphasizing the importance for *in-situ* characterization, especially for low-crosslinked particles. Additionally, the microgels' lateral size, estimated from phase images of deposited microgels, is typically lower than the one obtained directly from *in-situ* AFM at the liquid-liquid interface.

## Effect of temperature

pNIPAM microgels are most typically associated with their sharp temperature response in bulk aqueous conditions and the interplay between interfacial adsorption and temperature has also been extensively explored.[28,46–48] However, a direct insight on the conformation changes at the interface for temperatures below and above the solubility transition of pNIPAM is currently lacking.

Our AFM liquid cell allows for accurately controlling the temperature of the sample, enabling us to image the 3D conformation of the microgels across their volume phase transition temperature (VPTT). Figures 4d-f report the height profiles in the water and oil phase for each of the investigated microgels, below and above the VPTT, at 25 and 40°C, respectively. In all cases, at high temperature the particles are stretched out on the interface plane and assume a highly non-spherical shape, with $D_i \gg h_w+h_o$, maintaining a core-corona structure. In particular, the profiles on the oil side remain essentially unaltered (orange and red curves): hexadecane is a bad solvent for pNIPAM irrespective of the solution temperature and the portion of the microgels exposed to the oil is always collapsed. Conversely, a marked temperature dependence is seen for the portion of the particle in contact with water (light and dark blue curves), with substantial de-swelling upon crossing the VPTT.

Because the particles are confined at the interface, the de-swelling is anisotropic, different to the isotropic shrinkage of microgels in suspension. We quantify the extent of the interfacial volumetric swelling as a function of temperature in the water phase as the ratio between the volume occupied by the particles in water at 25° and 40°C. For both *CC5* and *CC1*, the volumetric swelling at the interface is much lower than the one in the bulk, reaching $3.0 \pm 0.19$ and $1.86 \pm 0.06$ for *CC5* and *CC1*, respectively, while their bulk volumetric swelling is respectively $10.6 \pm 0.3$ and $15.0 \pm 0.3$ (see Tables S1-S2). This indicates that, while the particles maintain a thermal responsiveness, the overall degree of deswelling is restrained by the fluid interface. These results corroborate literature data which reported the presence of a core-corona structure for standard microgels also above the VPTT, as evidenced by *ex-situ* AFM imaging,[48] as well as a decrease of the out-of-plane extension of the microgels into the water phase as measured by ellipsometry.[28,47]

As previously discussed, the *INV* microgel presents qualitative differences and its swelling behavior at the interface as a function of temperature emphasizes how the internal architecture controls the particle conformation and response to external stimuli. The presence of an ultra-low crosslinked core, which remains highly swollen in water at 25°, causes a pronounced conformational change when the solution temperature is increased above the VPTT (Fig. 4f). The entire polymer network in the water phase is now collapsed, up to an interfacial volumetric swelling of about 100, leaving only a very thin polymer layer on the fluid surface.

# DISCUSSION

The results reported in this work constitute a step forward in accessing the detailed conformation of responsive soft particles adsorbed at fluid interfaces. As it has been demonstrated in the case of bulk microgel systems, novel developments on the visualization of the microgels' internal network are of crucial importance to characterize such complex objects,[49,50] where insight into their 3D shape and deformation enabled an improved understanding of their phase diagram and rheological properties as a function of the effective particle concentration.[51–55] We believe that the imaging technique presented here will, similarly, enable advancing our understanding of the structural and mechanical properties of soft particle monolayers.

As an example, information on the conformation of adsorbed microgels across the VPTT can shed light on the mechanism behind the destabilization of microgel-stabilized emulsions by temperature increase.[46,47] Recently, it has been argued that the collapse of polymer chains onto the microgel core in the water phase plays a central role in causing emulsion destabilization at high temperature due to a decrease of the steric repulsion between two microgel-covered emulsion drops.[48] The findings we report here enable visualizing and quantifying such an effect, indicating that indeed pronounced deswelling on the water side takes place at high temperatures. Additionally, *in-situ* AFM imaging shows that the internal architecture of the microgels play a crucial role in controlling the particle volumetric swelling in the aqueous phase, suggesting that particles with a more loosely crosslinked core will perform as better stabilizers for the production of temperature-sensitive emulsions. Directly verifying how the particle cross-linking density profile unravels at the interface is thus of particular importance in studies on Pickering emulsions stabilization.[22,43]

Similarly, these findings illustrate that reconstructing the full 3D shape of the microgels is important to describe interactions in ordered monolayers, where interparticle contacts can happen both at the interface and through the bulk liquids, therefore opening the way to a more advanced control over the monolayer microstructure and mechanical properties in response to interfacial stresses.[33]

Reliable *ex-situ* imaging may not be possible for all oil phases, as for instance in the case of 1-decanol. Here, *in-situ* visualization of the microgel's conformation at the 1-decanol-water interface unambiguously shows how the particles deviate from the common "fried-egg" shape when the polymer solubility in the top phase increases, and the in-plane force exerted by interfacial tension decreases. Direct imaging illustrates that particle properties can also be tuned by changing the top fluid phase, in addition to modifying microgel architecture during synthesis.

Overall, we envisage that *in-situ* AFM imaging will greatly enhance the toolbox of available characterization techniques of microgel monolayers, which can now be applied to a multitude of soft particles at interfaces as an exciting way to explore their properties and tackle open questions.

## MATERIALS AND METHODS

### Reagents

N,N'-Methylenebis(acrylamide) (BIS, Fluka, 99.0%), methacrylic acid (MAA, Acros Organics, 99.5%), potassium persulfate (KPS, Sigma-Aldrich, 99.0%), isopropanol (Fisher Chemical, 99.97%), toluene (Fluka Analytical, 99.7%), n-hexane (SigmaAldrich, HPLC grade 95%), n-hexadecane (Acros Organics 99.0%) and 1-decanol (Sigma-Aldrich, ≥98%) were used without further purification. N-isopropylacrylamide (NIPAM, TCI 98.0%) was purified by recrystallization in 40/60 v/v toluene/hexane.

### Microgels synthesis

The microgels used in this study were synthesized by free-radical precipitation polymerization.

*Soft microgels - CC1.* NIPAM (0.385 g), 5 mol % MAA and 1 mol % BIS were dissolved in 25 mL of MQ water at room temperature. The reaction mixture was then immersed into an oil bath at 80°C and purged with nitrogen for 1 h. The reaction was started by adding 10 mg of KPS previously dissolved in 1 mL MQ water and purged with nitrogen. The polymerization was carried out for 6 h in a sealed flask. Afterwards, the colloidal suspension was cleaned by dialysis for a week, and 8 centrifugation cycles and resuspension of the sedimented particles in pure water.

*Stiff microgels - CC5.* NIPAM (1 g), 5 mol % MAA and 5 mol % BIS were dissolved in 50 mL of MQ water at room temperature. The reaction mixture was then purged with nitrogen for 1 h. Afterwards, 40 mL of the monomer solution was taken out with a syringe. 10 mL of MQ water were added to the reaction flask and the solution was immersed into an oil bath at 80°C and purged with nitrogen for another 30 min. The reaction was started by adding 13 mg of KPS previously dissolved in 1 mL MQ water and purged with nitrogen. After 1.5 minutes the solution turned slightly milky, and feeding of the monomer solution (40 mL at 0.5 $mL \cdot min^{-1}$) to the reaction flask was started. When feeding was terminated, the reaction was immediately quenched by opening the flask to let the air in, and placing it in an ice bath. The obtained colloidal suspension was cleaned by dialysis for a week, and by 8 centrifugation cycles and resuspension of the sedimented particles in pure water.

*Inverse microgels - INV*. This is a two-step synthesis devised to produce core-shell microgels having an ultra-low crosslinked core[56] covered by a crosslinked shell. NIPAM (0.5 g) and 5 mol % MAA were dissolved in 50 mL of MQ water at room temperature. The reaction mixture was then purged with nitrogen for 1 h and immersed into an oil bath at 80°C. 50 mg of KPS, previously dissolved in 2 mL MQ water and purged with nitrogen, were added to the flask to start the reaction. Meanwhile, in a separate flask a second monomers solution was prepared, containing NIPAM (0.5 g), 5 mol % MAA and 5 mol % BIS dissolved in 40 mL of MQ water, purged with nitrogen for 1 h, and then transferred in a syringe. Additionally, 13 mg of KPS were dissolved in 1 mL MQ water and purged with nitrogen. After 1 h 30 min since the beginning of the reaction, 13 mg of KPS were added to the reaction flask, immediately followed by the second monomers solution, which was added drop-wise at a feeding rate of 0.5 $mL \cdot min^{-1}$. When feeding was terminated, the reaction was immediately quenched by opening the flask to let the air in, and placing it in an ice bath. The obtained colloidal suspension was cleaned by dialysis for a week, and by 8 centrifugation cycles and resuspension of the sedimented particles in pure water.

## Methods

*DLS and SLS*. Dynamic light scattering (DLS) experiments were performed using a Zetasizer (Malvern, UK). The scattering vector for DLS experiments was q = 0.026 $nm^{-1}$. The samples were let to equilibrate for 15 min at the required temperature (22 or 40°C) prior to performing six consecutive measurements. For static light scattering (SLS), a CGS-3 Compact Goniometer (ALV, Germany) system was used, equipped with a Nd-YAG laser, $\lambda = 532$ nm, output power 50 mW before optical insulator, measuring angles from 30° to 150° with 5 or 2° steps. Static scattering form factor analysis was performed using the FitIt! tool developed by Otto Virtanen for MATLAB.[56] A detailed description of the fitting procedure is reported elsewhere.[33]

*Deposition of isolated microgels from a liquid-liquid interface.* Microgels were deposited from a hexane-water interface onto silicon wafers for atomic force microscopy (AFM) imaging of isolated dried particles following an already reported procedure.[33,42] Silicon wafers were cut into pieces and cleaned by 15 min ultrasonication in toluene, isopropanol, acetone, ethanol and MQ water. A piece was then placed inside a Teflon beaker on the arm of a linear motion driver and immersed in water. Successively, a liquid interface was created between water and n-hexane. Around 100 $\mu$L of the microgels suspension was injected at the interface after appropriate dilution in a 4:1 MQ-water:IPA solution. After 10 min equilibration time, extraction of the substrate was conducted at a speed of 25 $\mu$m·s$^{-1}$ to collect the microgels adsorbed at the liquid interface.

*AFM Imaging and Analysis.* Imaging of microgels at liquid-liquid interface was carried out by using Bruker Dimension Icon AFM. At first, a small well was made by applying a drop of UV curable glue (Norland Optical Adhesive 81 (NOA81) on a piece of silicon wafer (Si-Mat, Landsberg, Germany) by using a pipette tip. This well (average depth of 2 - 10 $\mu m$) acts as a reservoir for containing the subphase (oil or water). Fig. S1 in the supporting information shows a profilometer image of such a reservoir on the silicon surface. The wafer was then glued to a bio heater cell (MFP 3D, Asylum research, Oxford instrument). Before each AFM experiments, the cell was cleaned with ethanol and the silicon wafer was plasma cleaned for 10 s using a plasma pen (Piezobrush® PZ2, Reylon Plasma GMBH, Germany). For imaging from the oil phase, the reservoir was filled with 5 $\mu L$ of the microgel suspension in water. After 5 min, the entire cell was filled with the oil (hexadecane or 1-decanol). For imaging from the water phase, the reservoir was first filled with oil, then $\sim 5\mu L$ of the microgel suspension in water was injected on top of the oil. After 5 min, the entire cell was filled with water. The water was exchanged two times to avoid multilayer formation and to remove any excess of microgels floating in the bulk phase. The bio heater cell was placed under the AFM and the imaging was started after around 30 min to allow for the stabilization of the interface.

AFM imaging at the interface was carried out by using PeakForce tapping mode. For the hexadecane-water interface, cantilevers with a nominal spring constant of $\sim 0.12 N \cdot m^{-1}$ (PEAKFORCE-HIRS-F-B, Bruker) were chosen for imaging. The tip was approached to the interface by setting a PeakForce set point of 100 pN, and adjusted slightly along with the feedback gains once the tip was engaged at the interface. The PeakForce during the imaging was varied between 100 pN - 500 pN with the aim of obtaining images with the highest quality. The imaging at the 1-decanol-water interface was done by using much softer cantilevers due to the low interfacial tension (nominal spring constant of $\sim 0.03 N \cdot m^{-1}$, CSG01, NT-MDT). The PeakForce set point during the engaging was kept as low as 5 pN and was varied between 5 - 20 pN while imaging in order to avoid snapping-in of the cantilever into the subphase. The PeakForce amplitude during imaging in the various fluid phases was varied between 100 - 300 nm. The oscillation frequency was chosen between 1 - 2 KHz, and the scanning speed between 0.2 - 1 Hz. Along with topographical images, adhesion and deformation images were also captured in PeakForce tapping mode.

Dry microgels deposited on silicon wafers were imaged in tapping mode, using cantilevers with ∼300 kHz resonance frequency and ∼ $26 mN \cdot m^{-1}$ spring constant (AC160TS-R3, Olympus cantilevers, Japan). Height and phase images were recorded at the same time.

All AFM images were first processed with open-source software Gwyddion and successively analysed with custom MATLAB codes. Imaging from the water side is subjected to more noise with respect to imaging from the oil phase, especially between lines perpendicular to the scanning direction. Therefore, some images, such as the one in Fig. 1d, have been corrected with a correlation averaging algorithm in the Gwyddion software prior to further analysis. The following procedure was used to obtain an averaged height profile: for each microgel, horizontal and vertical profiles passing through its center were extracted. Subsequently, an average over at least 10 microgels was obtained by aligning each profile by its center value. To reconstruct the entire profile of a microgel adsorbed at the fluid interface, the profile measured on the water side was inverted to appear below the interface plane. The 3D reconstructions in Fig. 1g and 3d are obtained by rotating the height profiles for $r > 0$ around the y-axis. The volume occupied by the particle in the water phase was calculated as:

$$V_{int} = \pi \int [f(r)]^2 \, dr$$

where $f(r)$ is the radial profile in the water phase at the given temperature.

*Profilometry.* A 3D optical profiler (Sensofar PLu Neox, Sensofar-Tech, SL., Terrassa, Spain) operating in confocal mode with a 5× objective was used to measure the depth of the reservoir made on the silicon wafer.

# Acknowledgments

The authors acknowledge Dr. Kirill Feldman and Prof. Jan Vermant for access to SLS measurements, and for useful discussions. J.V. acknowledges funding from the European Union's Horizon 2020 research and innovation programme under the Marie Skłodowska Curie grant agreement 888076.

# Authors Contributions

Author contributions are defined based on the CRediT (Contributor Roles Taxonomy) and listed alphabetically. Conceptualization: Formal Analysis: Funding acquisition: Investigation: Methodology: Project Administration: Software: Supervision: Validation: Visualization: Writing - original draft: Writing - review and editing: .

# Supplementary Material

# Supplementary Experimental Data

Table S1: Microgels hydrodynamic diameter ($D_h$) in aqueous solution

| Microgel | $D_h$ at 22°C [nm] | $D_h$ at 40°C [nm] | Volumetric swelling ratio |
|---|---|---|---|
| 5 mol % BIS | 1150 ± 27 | 525 ± 3 | 10.6 ± 0.3 |
| 1 mol % BIS | 786 ± 13 | 319 ± 1 | 15.0 ± 0.3 |
| "inverse" | 903 ± 16 | 324 ± 0.4 | 21.6 ± 0.4 |

Table S2: Microgels interfacial dimensions[a] at 25°C and volumetric swelling ratio

| Microgel | $D_i$ [nm] | $h_w$ [nm] | $h_o$ [nm] | Interfacial volumetric swelling ratio[b] |
|---|---|---|---|---|
| 5 mol % BIS | 2088 ± 250 | 488 ± 34 | 71 ± 5 | 3.0 ± 0.19 |
| 1 mol % BIS | 1556 ± 180 | 295 ± 15 | 21 ± 1 | 1.86 ± 0.06 |
| "inverse" | 2160 ± 270 | 396 ± 37 | 12 ± 2 | 144 ± 50 |

[a] $D_i$ is the particle diameter at the interface as measured by AFM height images from the water side. $h_w$ and $h_o$ are the maximum particle height in water and oil, respectively. [b] Swelling ratio of microgels adsorbed at the fluid interface, calculated as the volumetric ratio between the volume occupied by the particles in water at 25°C and 40°C.

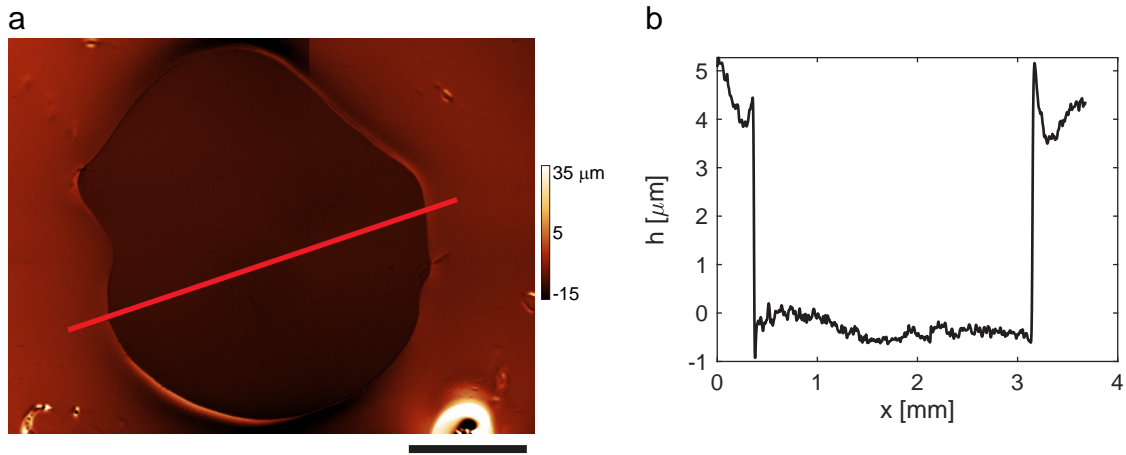

Figure S1: **3D image of the experimental cell** a) Optical profilometer image and b) height profile along the red line in (a) of a reservoir made of a thin ring of UV-curable glue on a silicon wafer. Scale bar: 1 mm.

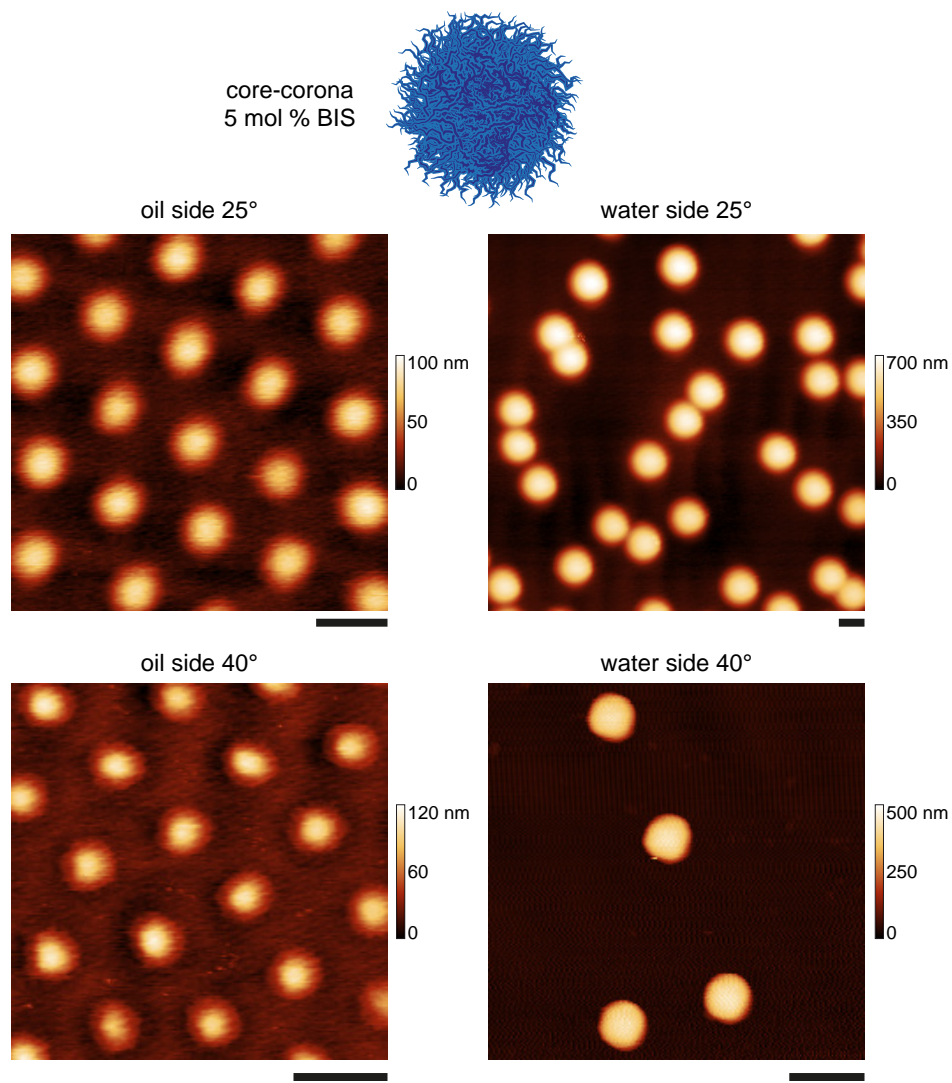

Figure S2: **AFM profiles of *CC5* microgels at the hexadecane-water interface** AFM height images taken at the fluid interface, at 25 (top row) and 40°C (bottom row), with tip in the oil (left column) and water phase (right column). Scale bars: 1 $\mu$m.

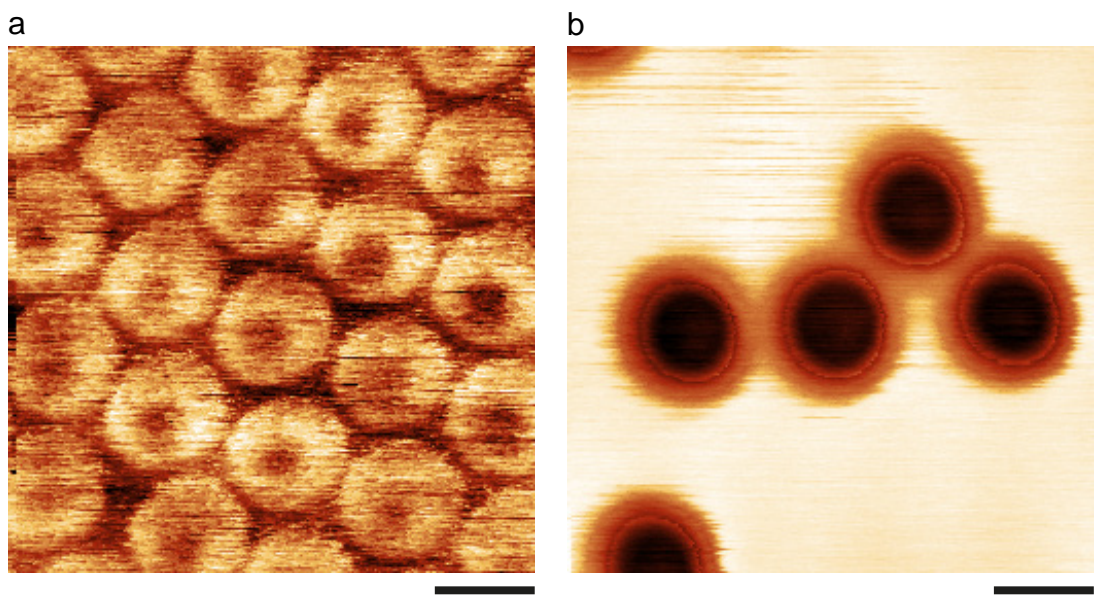

Figure S3: **AFM adhesion images at the hexadecane-water interface.** AFM adhesion images captured from the oil (a) and water (b) side, for *CC5* microgels. The image in (a) corresponds to the height image in Fig. 2a; the image in (b) to the height image in Fig. 2b. Scale bars: 1 $\mu$m.

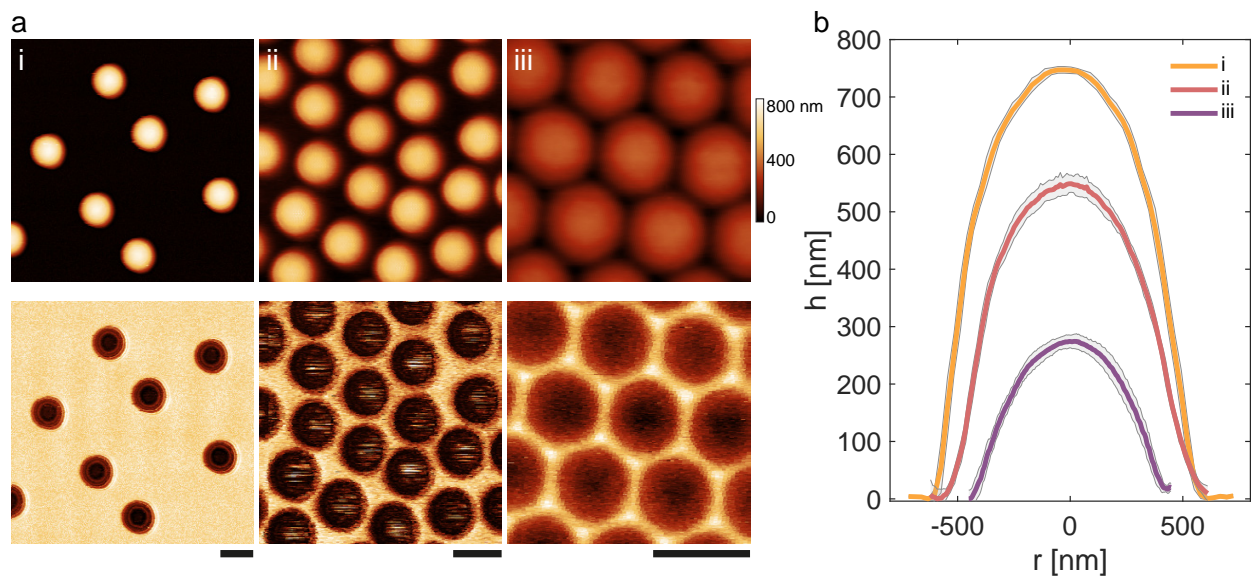

Figure S4: **AFM images at the 1-decanol-water interface as a function of microgel density.** (a) AFM height (top) and adhesion (bottom) images of *CC5* microgels imaged from the 1-decanol side. The microgel interfacial concentration is increased from image (i) to (iii). Scale bars: 1 $\mu$m. (b) Averaged height profiles of the microgels in (a). For images (ii) and (iii), $h = 0$ is the lower height measured by the AFM tip, and does not correspond to the fluid interface.

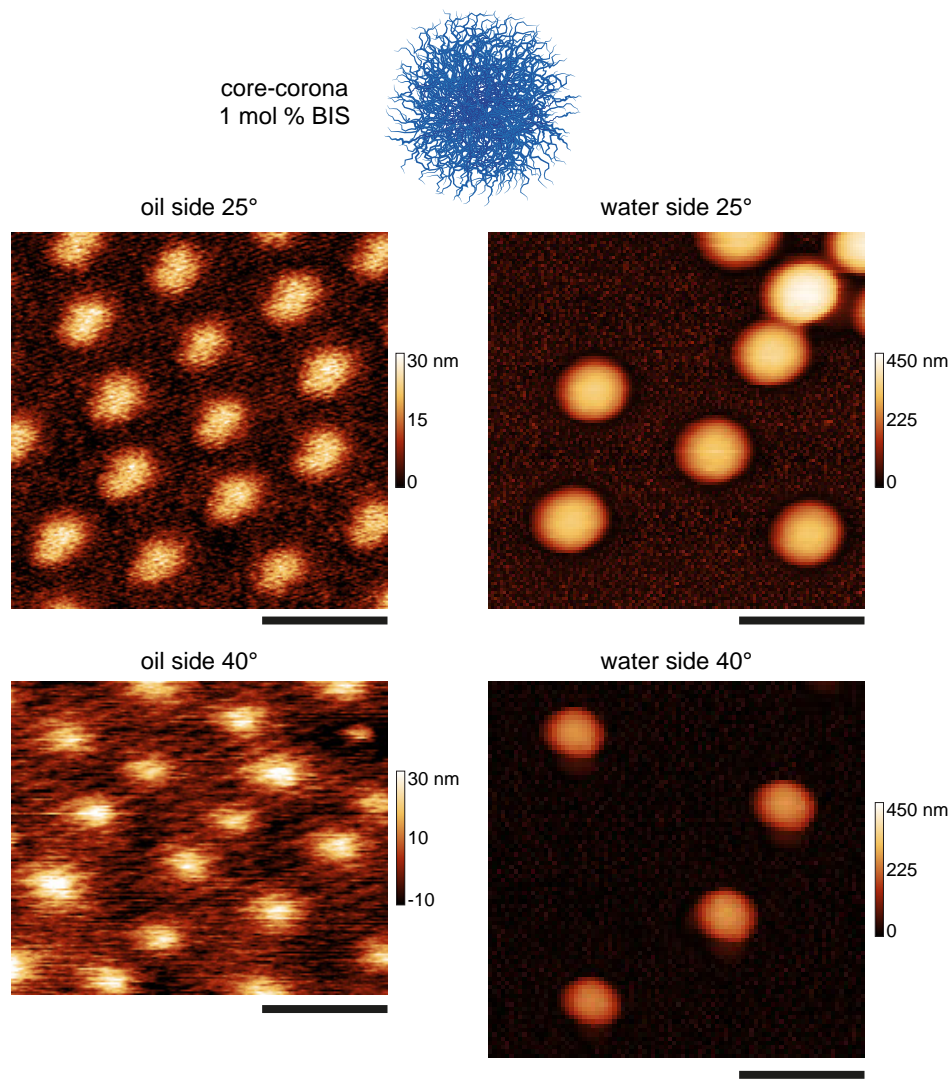

Figure S5: **AFM profiles of *CC1* microgels at the hexadecane-water interface** AFM height images taken at the fluid interface, at 25 (top row) and 40°C (bottom row), with tip in the oil (left column) and water phase (right column). Scale bars: 1 $\mu$m.

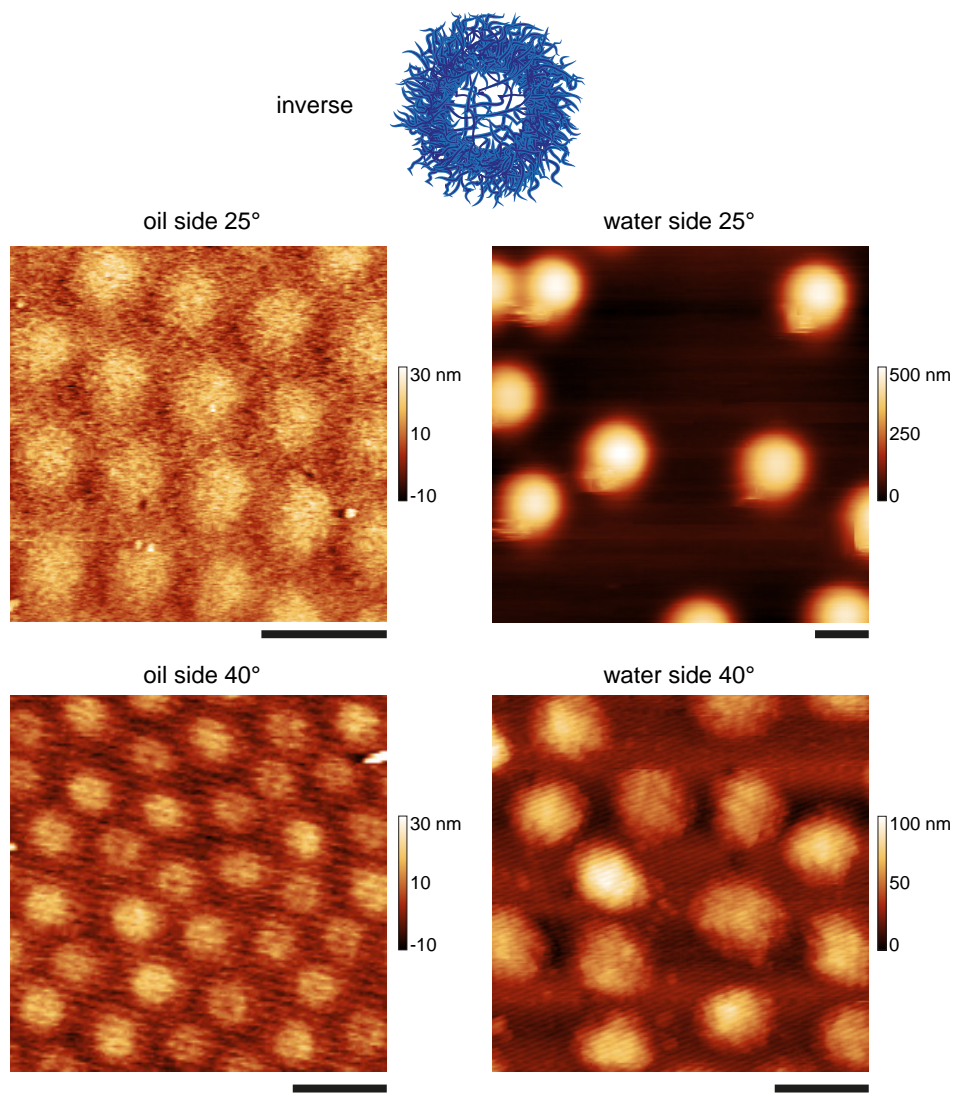

Figure S6: **AFM profiles of *INV* microgels at the hexadecane-water interface** AFM height images taken at the fluid interface, at 25 (top row) and 40°C (bottom row), with tip in the oil (left column) and water phase (right column). Scale bars: 1 $\mu$m.

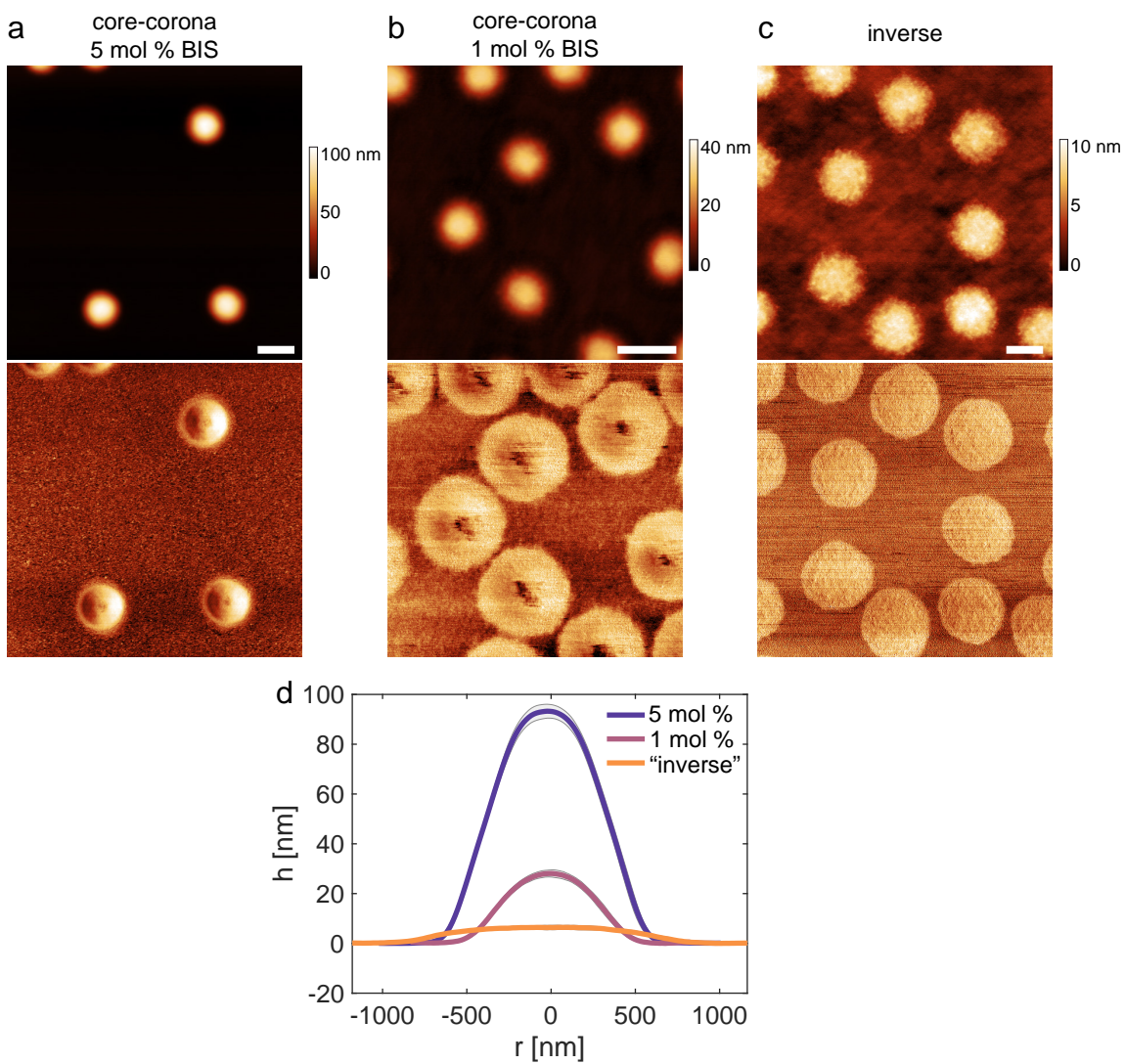

Figure S7: **AFM profiles of dried microgels** (a-c) Representative AFM height (top) and phase (bottom) images of microgels transferred from the hexane-water interface onto a silicon wafer, and imaged in dry condition. Scale bar: 1 µm. d) Experimental height profiles in dry condition. The shaded regions correspond to the standard deviations of the height profiles calculated on around 20 particles.

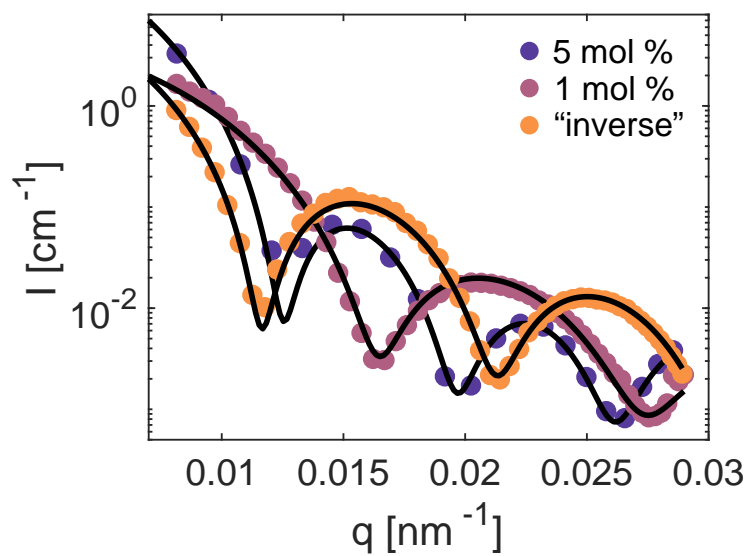

Figure S8: **Microgel form factors in aqueous solution.** Experimental form factors obtained from SLS experiments at 25°C. Black lines are fits (see Methods).

## Force curves at the fluid interface

The PeakForce tapping technique enables the possibility of simultaneously acquiring force *vs* distance curves along with the topographical images. By knowing the spring constant of the cantilever and the deflection sensitivity values, one can obtain the force as a function of the tip-to-sample separation. Figure S9 shows the extracted force *vs* separation curves from the PeakForce topographical image obtained for the sample *CC5* at the hexadecane-water interface.

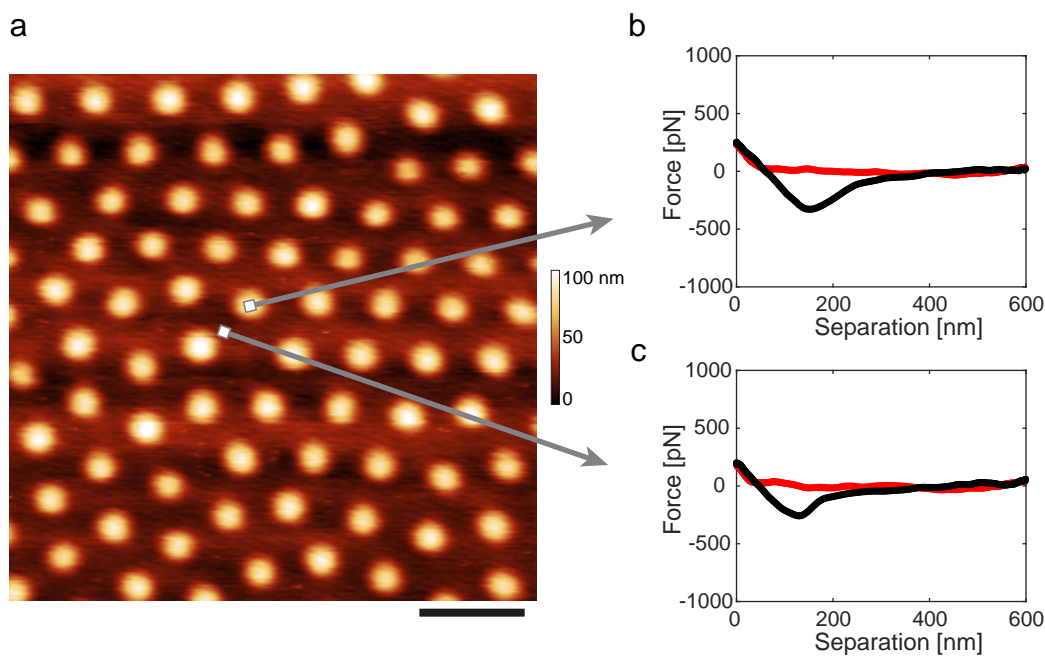

Figure S9: **PeakForce curves at the fluid interface** a) PeakForce image of microgel *CC5* at the hexadecane-water interface, imaged from the oil side. b-c) Approach (red) and retract (black) force curves from the image in (a), measured on a microgel (b) and in-between two microgels (c). The slope from the retraction curve for b) and c) was found to be 4.3 and 4.5 mN/m, respectively. Scale bar: 2 $\mu$m.

The slope of the force *vs* separation curves taken at the hexadecane-water interface can be used to estimate its interfacial tension, which is found to be of around 4 mN/m. Comparable estimations, around 15 mN/m, were reported for the octane-water interface, by using a similar AFM imaging technique.[39] Such low values of the measured interfacial stiffness with respect to typical values obtained when imaging a solid substrate, can be used as a direct confirmation that the images are taken at a fluid interface, and that there was no influence of the underlying substrate during the PeakForce imaging.